\documentclass[aps,twocolumn,groupedaddress,showpacs,prb,floatfix]{revtex4}
\usepackage{graphicx,rotating,subfigure,amsmath,amsfonts,amssymb,delarray}
\renewcommand{\vec}[1]{\boldsymbol #1}
\newcommand{\e}{\text{e}}

\def\wt{\widetilde}

\begin{document}
\bibliographystyle{apsrev}


\title{Quantum versus classical behavior in the boundary susceptibility of the ferromagnetic Heisenberg chain}


\author{J.~Sirker}
\email[]{sirker@physics.ubc.ca}
\affiliation{Department of Physics and Astronomy, University of British Columbia,
Vancouver, B.C., Canada V6T 1Z1}
\author{M.~Bortz}
\affiliation{Department of Theoretical Physics, Research School of Physics and Engineering, Australian National University, Canberra ACT 0200, Australia}

\date{\today}

\begin{abstract}
  We calculate the temperature dependence of the boundary susceptibility
  $\chi_B$ for the quantum ferromagnetic Heisenberg chain by a modified
  spin-wave theory (MSWT). We find that $\chi_B$ diverges at low temperatures
  $\sim -T^{-3}$ and therefore more rapidly and with opposite sign than the
  bulk susceptibility $\chi_{\text{bulk}}\sim T^{-2}$. Our result for $\chi_B$
  is identical in leading order with the result for the classical system. In
  next leading orders, however, quantum corrections to the classical result
  exist which are important to obtain a good description over a wide
  temperature range. For the $S=1/2$ case, we show that our full result from
  MSWT is in excellent agreement with numerical data obtained by the
  density-matrix renormalization group applied to transfer matrices. Finally,
  we discuss the quantum to classical crossover as well as consequences of our
  results for experiment in some detail.
\end{abstract}

\maketitle

\section{Introduction}
\label{Intro}
Although the classical and the quantum version of the ferromagnetic
Heisenberg model have the same ground state, it is not obvious in how
far the low temperature properties of these systems are also
similar. The Hamiltonian for the quantum ferromagnetic chain with open
boundary conditions (OBC), spin $S$ and $N$ sites in a magnetic field $h$ along
the $z$-direction is given by
\begin{equation}
\label{eq_i1}
H = -J\sum_{n=1}^{N-1} \vec{S}_n\vec{S}_{n+1} -g\sum_{n=1}^N h S^z_n
\end{equation}
where $J>0$ is the coupling constant and $g$ the g-factor. The
classical version of this model can be obtained by introducing new
unit vector operators $\vec{s}_n = \vec{S}_n/S$. \cite{Fisher} These
new operators commute in the limit $S\rightarrow\infty$, leading to the
classical Hamiltonian
\begin{equation}
\label{eq_i2}
H = -J_c\sum_{n=1}^{N-1} \vec{s}_n\vec{s}_{n+1} -g_c\sum_{n=1}^N h s^z_n \; .
\end{equation}
To allow for a comparison between the quantum and the classical Hamiltonian
for different spin values $S$ we have set $J_c = J S^2$ and $g_c = g S$.

The finite temperature properties of the classical model (\ref{eq_i2})
have been calculated several decades ago for OBC \cite{Fisher} as well
as periodic boundary conditions (PBC) \cite{Joyce}. Whereas the first
correction to the total free energy in the thermodynamic limit for PBC
is $\mathcal{O}(1/N)$, the free energy for OBC contains a term
$\mathcal{O}(1)$. This boundary or surface free energy then yields
$\mathcal{O}(1)$-contributions to all other thermodynamic quantities
as for example the susceptibility. From Fisher's results \cite{Fisher}
one finds that the classical bulk susceptibility behaves as
\begin{equation}
\label{eq_i3}
\chi^c_{\text{bulk}} = N g_c^2\left(\frac{2J_c}{3T^2} - \frac{1}{3T}\right)
\end{equation}
whereas the classical boundary susceptibility ($\mathcal{O}(1)$
contribution) is given by
\begin{equation}
\label{eq_i4}
\chi^c_B = -g_c^2\left(\frac{2J_c^2}{3T^3} - \frac{4J_c}{3T^2} + \frac{1}{3T}\right) \; .
\end{equation}
$\chi^c_B$ therefore diverges more rapidly and with opposite sign than
$\chi^c_{\text{bulk}}$. 

A very different behavior for bulk and boundary susceptibility has also
recently been observed for the quantum antiferromagnetic $S=1/2$ XXZ-chain
with anisotropy $0\leq \Delta \leq 1$.
\cite{FujimotoEggert,FurusakiHikihara,BortzSirker} For this system it is known
that the bulk susceptibility is finite for $T\rightarrow 0$ with the $T=0$
value of $\chi_{\text{bulk}}$ depending on the anisotropy $\Delta$.
$\chi_B(T)$, on the other hand, is finite only for $0\leq\Delta < 1/2$ whereas
it diverges for $1/2\leq\Delta \leq 1$ when $T\rightarrow 0$. By a combination
of different techniques like bosonization, conformal field theory, Bethe
ansatz as well as numerical results, a complete picture of the low-temperature
properties of $\chi_B$ has been obtained.
\cite{FujimotoEggert,FurusakiHikihara,BortzSirker} These results are not only
of theoretical interest but might also be relevant for realizations of quasi
one-dimensional antiferromagnets as for example SrCuO$_3$, in particular, when
such compounds are doped with a moderate amount of non-magnetic impurities. In
such a case the spin chain will be partitioned into finite chains with
essentially free boundaries and knowledge of the boundary contributions will
be essential to understand experiments on such systems.

For the quantum $S=1/2$ ferromagnetic chain the standard bosonization approach
and conformal field theory are not applicable because the dispersion relation
is quadratic instead of linear. The model is, however, still integrable and
thermodynamic properties can in principle be calculated either by the
thermodynamic Bethe ansatz \cite{TakahashiMSWT} (TBA) or by the Bethe ansatz
applied to quantum transfer matrices.\cite{JungerIhle} The bulk susceptibility
has indeed been obtained by an analysis of the TBA equations. In the first
analysis of this kind by Schlottmann \cite{Schlottmann} it has been proposed
that $\chi_{\text{bulk}}\sim J/T^2\ln(J/T)$. A later numerical analysis of the
TBA \cite{TakahashiMSWT,TakahashiMSWT2,TakahashiMSWT3}, however, showed that
$\chi_{\text{bulk}}\sim J/6T^2$ at low temperatures as in the classical case
but with corrections to this leading term which are different from
Eq.~(\ref{eq_i3}). In addition it has been found that the leading term as well
as the quantum corrections can be obtained by a modified spin-wave theory
(MSWT). \cite{TakahashiMSWT,TakahashiMSWT4} It has later been shown that the classical and
quantum ferromagnetic chains obey the same scaling laws at low temperatures.
\cite{NakamuraHatano} Furthermore, the critical theory controlling the
low-energy behavior of both chains has been identified, which explains in more
general terms why $\chi^c_{\text{bulk}}$ and $\chi_{\text{bulk}}$ are
identical at low temperatures.  \cite{TakahashiNakamura}

It is still unclear how the TBA has to be modified to allow also for the
calculation of boundary contributions. Some of the difficulties one encounters
are discussed in Refs.~\onlinecite{BortzSirker,GoehmannBortz}. Within the QTM
approach an explicit formula for the boundary free energy has been derived
very recently. \cite{GoehmannBortz} The explicit evaluation of this formula,
however, is still a formidable task because it involves expectation values of
an operator in the dominant eigenstate of the QTM which are notoriously
difficult to calculate.

For these reasons we will follow here a different route and will use
in section \ref{MSWT} Takahashi's MSWT, which has been so successful
for the bulk, to calculate the boundary susceptibility. In section
\ref{TMRG} we then compare our result with numerical data obtained by
the density-matrix renormalization group applied to transfer matrices
(TMRG). In the last section we discuss the quantum to classical
crossover observed and comment on the relevance of our results for
experiment.
\section{Modified spin-wave theory}
\label{MSWT}
With the help of the Holstein-Primakoff transformation 
\begin{equation}
\label{eq_MSWT1}
S_n^+ = \sqrt{2S}\sqrt{1-a^\dagger_n a_n/2S}\, a_n \quad , \quad S_n^z =S-a^\dagger_n a_n
\end{equation}
the model (\ref{eq_i1}) can be represented exactly in terms of bosons $a_n$.
Linear spin-wave theory is obtained if one replaces the second square root in
Eq.~(\ref{eq_MSWT1}) by 1. Corrections to this simple approximation can be
calculated in principle in a systematic way by expanding the square root in
powers of $1/S$. In any of these approximations it is important to notice that
the bosons have to obey a hard-core constraint restricting the maximum number
of bosons per site to $2S$. In higher dimension it is often acceptable to
ignore this constraint completely. In one dimension, however, this constraint
is crucial but hard to incorporate locally. Because the $SU(2)$ symmetry in
a system with $h=0$ can only be broken at $T=0$ we might instead try to
introduce a constraint which fixes the number of bosons to be $S$ on the
average so that 
\begin{equation}
\label{constraint}
\frac{1}{N}\sum_n \langle S^z_n\rangle  = 0
\end{equation} 
at any finite temperature. This approach has been used successfully by
Takahashi \cite{TakahashiMSWT} to calculate the free energy and the
susceptibility for a chain with PBC. We will use the same approach here for a
system with OBC to obtain the boundary susceptibility. 

Let us first rederive Takahashi's result for PBC in a slightly different way.
Expanding up to quartic order in the boson operators in Eq.~(\ref{eq_MSWT1})
and using a one-loop approximation for the quartic terms, we obtain for the
Hamiltonian (\ref{eq_i1}) at zero magnetic field
\begin{eqnarray}
\label{eq_MSWT2}
H &=& J S' \sum_k \epsilon(k) a_k^\dagger a_k + v \sum_k a_k^\dagger a_k \nonumber \\
\epsilon(k) &=& 2(1-\cos k)  
\end{eqnarray}
where 
\begin{equation}
\label{eq_MSWT3}
S' = S -\frac{1}{2N}\sum_k \epsilon(k) n_k \; ,
\end{equation}
and the average number of bosons $n_k$ is given by
\begin{equation}
\label{eq_MSWT4}
n_k :=\langle a_k^\dagger a_k\rangle = [\exp(JS'\epsilon(k)/T+v)-1]^{-1}\; .
\end{equation}
At temperatures $T/J< 1$ the number of bosons in high momentum states is
small. The bosons in low momentum states, on the other hand, will yield only a
small contribution to the sum in Eq.~(\ref{eq_MSWT3}) so that we will set
$S'=S$ in the following. According to Eq.~(\ref{constraint}), the potential
$v$ then has to be determined in such a way that
\begin{equation}
\label{eq_MSWT5}
S=\frac{1}{N}\sum_k n_k \; .
\end{equation}
Differentiating the partition function for the Hamiltonian (\ref{eq_i1}) twice
with respect to $h$ one finds that the susceptibility is given by $\chi =
g^2/T \sum_{n,m} \langle S^z_n S^z_m\rangle$. However, the spin-wave expansion
we are using here breaks the $SU(2)$ symmetry so that we will calculate the
susceptibility instead by
\begin{equation}
\label{eq_MSWT6}
\chi = \frac{g^2}{3T} \left\{\sum_{n=1}^N\sum_{m\neq n} \langle \vec{S}_n\vec{S}_m\rangle +NS(S+1)\right\} \; .
\end{equation}
In this way the consequences of $SU(2)$ symmetry breaking are less
severe due to the averaging over all three directions. Using the
constraint (\ref{eq_MSWT5}) one finds
\cite{TakahashiMSWT}
 \begin{equation}
\label{eq_MSWT7}
\langle \vec{S}_n\vec{S}_m\rangle = \left(\frac{1}{N}\sum_k \cos[k(r_n-r_m)] n_k\right)^2 \: .
\end{equation} 
The momenta for a chain of length $N$ with PBC are given by $k=2\pi
l/N$ where $l=0,1,\cdots,N-1$. For $T/J\ll 1$ the most important
contributions to the sum in (\ref{eq_MSWT7}) come from $k\approx
0,2\pi$ and we can evaluate these contributions by using a saddle
point integration 
\begin{eqnarray}
\langle \vec{S}_n\vec{S}_m\rangle &\approx& \left(\frac{1}{2\pi}\int_0^{2\pi}\frac{\cos[k(r_n-r_m)]}{\e^{\beta JS\epsilon(k)+v}-1}dk\right)^2\label{int}\\
&\approx& \left(\frac{T}{JS\pi}\int_0^\infty \frac{\cos[k(r_n-r_m)]}{k^2 +Tv/JS}dk\right)^2\nonumber \\
 &=& \frac{t}{4v}\exp(-2\sqrt{tv}|r_n-r_m|) \label{eq_MSWT8}\: .
\end{eqnarray} 
In the last line we have introduced the abbreviation $t=T/JS$. To understand
why the saddle point approximation for the integrand in (\ref{int}) is indeed
sufficient here we make the following observation: (\ref{int}) can be
evaluated alternatively by closing the integration contour in the upper or
lower half of the complex $k$-plane, depending on the sign of $r_n-r_m$. The
residues closest to the real axis then yield (\ref{eq_MSWT8}). Next-leading
residues give contributions $\mathcal O(\sqrt T\exp[-\sqrt{T}])$. These would
result in terms $\mathcal O(1/\sqrt{T})$ in the susceptibility, which are
neglected in the ongoing.

From the constraint (\ref{eq_MSWT5}) one can easily determine the potential
$v$ as a series in $\sqrt{t}$. The result is \cite{TakahashiMSWT}
 \begin{equation}
\label{eq_MSWT9}
\sqrt{v} =
\frac{\sqrt{t}}{2S}+q\left(\frac{\sqrt{t}}{2S}\right)^2+q^2\left(\frac{\sqrt{t}}{2S}\right)^3
+\mathcal{O}(t^2) \: ,
\end{equation} 
where $q= \zeta(1/2)/\sqrt{\pi}$. When we rewrite the correlation
function (\ref{eq_MSWT8}) in terms of the normalized spin operators
$\vec{s}_n$ and the coupling constant $J_c$ as given in the
introduction and use only the leading term from (\ref{eq_MSWT9}) we find
\begin{equation}
\label{eq_MSWT10}
\langle \vec{s}_n\vec{s}_m\rangle = \exp(-|r_n-r_m|T/J_c)
\end{equation} 
for all values of $S$. In particular, the correlation length at $T/J\ll 1$ is
always given by $\xi=J_c/T$. Furthermore, Eq.~(\ref{eq_MSWT10}) also agrees
with the result for the classical model. \cite{Fisher} Note, however, that
this is no longer the case if one takes the next-leading terms in
(\ref{eq_MSWT9}) into account.

To calculate the susceptibility we have to evaluate the sum in
Eq.~(\ref{eq_MSWT6}). For PBC each distance $|r_n-r_m|=1,\cdots,N/2$
appears $2N$ times. The susceptibility in the thermodynamic limit can
therefore be obtained by
\begin{eqnarray}
\label{eq_MSWT11}
\chi_{\text{\tiny PBC}}\!\! &=& \!\!\lim_{N\rightarrow\infty}\frac{g^2}{3T} \left\{\frac{Nt}{2v}\sum_{r=1}^{N/2}\e^{-2r\sqrt{tv}} +NS(S+1)\right\} \nonumber \\
&=& \frac{Ng^2}{12JS}\bigg(t^{-1/2}v^{-3/2}-v^{-1} \nonumber \\
&+& 4S(S+1)t^{-1}+\mathcal{O}(\e^{-N})\bigg) \; .
\end{eqnarray} 
The first term agrees exactly with the result obtained by Takahashi
\cite{TakahashiMSWT}, however, we find here in addition the second and third
term, which are absent in Takahashi's result. Note, that these terms exactly
cancel each other for $S\rightarrow\infty$. The differences between our and
Takahashi's result can be explained as follows: Whereas in
Ref.~\onlinecite{TakahashiMSWT} the sum in Eq.~(\ref{eq_MSWT6}) is carried out
{\it without} approximating the correlation function (\ref{eq_MSWT7}) we have
taken here only the long-distance asymptotics of $\langle
\vec{S}_n\vec{S}_m\rangle$ into account as obtained by the saddle point
approximation in (\ref{eq_MSWT8}). Interestingly, the terms in the
susceptibility up to $\mathcal{O}(1/T)$ remain unaffected, i.e., these terms
are not influenced by the behavior of the correlation function at
short-distances. In fact, we might trust our spine-wave approximation only in
the long-wavelength limit where the spin-wave interaction is small. In our
one-loop approximation this becomes clear when considering
Eqs.~(\ref{eq_MSWT2},\ref{eq_MSWT3}). When all momenta involved are small, the
sum in Eq.~(\ref{eq_MSWT3}) is also small and $S'\approx S$. In this limit the
Hamiltonian (\ref{eq_MSWT2}) becomes equivalent to the one for ideal
non-interacting spin waves.

For these reasons we cannot expect that the MSWT gives reasonable results if
we try to calculate local quantities for OBC near the boundary. We observed
that neither a local constraint $\langle S_n^z\rangle=0$ nor the correlation
function $\langle \vec S_n \vec S_m\rangle$ can be calculated without
inconsistencies.  For example, if we calculate the correlation function for
OBC explicitly we find a constant term which vanishes only if we set
$v=t/4S^2$ exactly. However, the condition (\ref{eq_MSWT5}) still requires
corrections to $v$ as given in (\ref{eq_MSWT9}).

Far enough away from the boundaries, on the other hand, the correlation
function will still behave as in Eq.~(\ref{eq_MSWT8}). When we perform the sum
in (\ref{eq_MSWT6}) using again this long-distance asymptotics for $\langle
\vec{S}_n\vec{S}_m\rangle$ but in a way appropriate for OBC we will already
obtain a $\mathcal{O}(1)$ correction to the susceptibility {\it without} taken
the modifications to the correlation function near the boundary into account.
We conjecture that for low temperatures, this term yields $\chi_B$. The
physical picture behind this procedure is as follows: We can combine two open
chains each of length $M-1$ to one periodic chain of length $N=2 M$, where the
two additional sites do not couple with their neighbors. We then carry out the
sum in Eq.~(\ref{eq_MSWT6}) only over one half of the periodic chain, thereby
discarding correlations between this subsystem and the rest. Doing so we
ignore local differences between PBC and OBC. 

What makes us confident that this is indeed sufficient to obtain the leading
terms in a low-temperature expansion for the boundary susceptibility $\chi_B$
is that the leading term $\sim -1/T^3$ is universal in the sense that it does
not depend on $S$. Especially, it is the leading term of $\chi_B$ for both
$S=1/2$ and $S=\infty$. \cite{NakamuraHatano,TakahashiNakamura} The classical
result for the correlation function (\ref{eq_MSWT10}) has been first obtained
by Fisher \cite{Fisher} for an {\it open chain}. I.e., in the classical limit
the exponential decay of the correlation function does depend only on
$|r_n-r_m|$ and not on $r_n,r_m$ alone, although translational invariance is
broken!

We therefore conjecture that the leading terms in a low-temperature
expansion of the susceptibility for a quantum chain with OBC are
given by
\begin{eqnarray}
\label{eq_MSWT12}
\chi_{\text{\tiny OBC}} &=& \frac{g^2}{3T} \left\{\frac{t}{4v}\sum_{\stackrel{n,m=1}{n\neq m}}^{N}\e^{-2|r_n-r_m|\sqrt{tv}} +NS(S+1)\right\} \nonumber \\
&=& \frac{Ng^2}{12JS}\bigg(t^{-1/2}v^{-3/2}-v^{-1} \\
&+& 4S(S+1)t^{-1}-\frac{1}{2N}t^{-1}v^{-2}+\mathcal{O}(\e^{-N})\bigg) \nonumber \; .
\end{eqnarray} 
In particular, the boundary susceptibility is given by
\begin{eqnarray}
\label{eq_MSWT13}
\chi_B &=& -\frac{g^2}{24JS}t^{-1}v^{-2} \nonumber \\ 
&=& -g^2\frac{2S^3}{3Jt^3}\left(1-q\frac{2\sqrt{t}}{S}+q^2\frac{3t}{2S^2}+\cdots\right) \: .
\end{eqnarray}
Note that the leading term is identical to the leading term in the
classical result (\ref{eq_i4}) when $J,g$ are replaced by
$J_c,g_c$. This confirms our expectations. To test if the procedure
proposed here gives indeed the right corrections to the classical
result we will check formula (\ref{eq_MSWT13}) against numerical data
for the $S=1/2$ quantum model in the following section.
\section{Numerical results} 
\label{TMRG} 
In a system with OBC the one-point correlation function $\langle
S^z(r)\rangle$ is no longer a constant because translational
invariance is broken. We define
\begin{equation}
\label{eq_FT1}
C(r) =  \langle S^z(r)\rangle_{\mbox{\footnotesize OBC}} - m
\end{equation}
where $m$ is the magnetization per site in the system with PBC and $r$ is the
distance from the boundary. The local boundary susceptibility is then given by
$\chi_B(r) = \partial C(r)/\partial h|_{h=0}$ and the total boundary
susceptibility $\chi_B$ can be obtained by
\begin{equation}
\label{eq_FT3}
\chi_B = \sum_{r=1}^{\infty} \chi_B(r) = \chi_{\mbox{\footnotesize OBC}} -
\chi_{\mbox{\footnotesize PBC}} \; .
\end{equation}
This means that we can calculate $\chi_B$ by considering only a local quantity
which is particularly useful in numerical calculations where it is difficult
to obtain the $\mathcal{O}(1)$ contribution directly. Particularly suited for
this purpose is the density-matrix renormalization group applied to transfer
matrices (TMRG) because the thermodynamic limit is performed exactly. The idea
of the TMRG is to express the partition function $Z$ of a one-dimensional
quantum model by that of an equivalent two-dimensional classical model which
can be derived by the Trotter-Suzuki formula. \cite{Trotter,Suzuki2} For the
classical model a suitable transfer matrix $T$ can be defined which allows for
the calculation of all thermodynamic quantities in the thermodynamic limit by
considering solely the largest eigenvalue of this transfer matrix. Details of
the algorithm can be found in
Refs.~\onlinecite{BursillXiang,WangXiang,Shibata,SirkerKluemperEPL}. The
method has been extended to impurity problems in Ref.~\onlinecite{RommerEggert}. In
particular, the local magnetization at a distance $r$ from the boundary of a
system with $N$ sites can be obtained by
\begin{equation}
\label{eq_FT4}
\langle S^z(r)\rangle = \frac{\sum_n
  \langle\Psi_L^n|T(S^z)T^{r-1}\wt{T}T^{N-r-1}|\Psi_R^n\rangle}
{\sum_n \langle\Psi_L^n|T^{N-1}\wt{T}|\Psi_R^n\rangle}
\end{equation}
where $|\Psi_R^n\rangle$ ($\langle\Psi_L^n|$) are the right (left) eigenstates
of the transfer matrix $T$, $\wt{T}$ is a modified transfer matrix
containing the broken bond and $T(S^z)$ is the transfer matrix with the
operator $S^z$ included. Because the spectrum of $T$ has a gap between the
leading eigenvalue $\Lambda_0$ and the next-leading eigenvalues,
Eq.~(\ref{eq_FT4}) reduces in the thermodynamic limit to
\begin{equation}
\label{eq_FT5}
\lim_{N\rightarrow\infty}\langle S^z(r)\rangle = \frac{\langle\Psi_L^0|T(S^z)T^{r-1}\wt{T}|\Psi_R^0\rangle}
{\Lambda_0^r \langle\Psi_L^0|\wt{T}|\Psi_R^0\rangle} \; .
\end{equation}
Therefore only the leading eigenvalue and the corresponding eigenvectors have
to be known to calculate the local magnetization in the thermodynamic limit.
Far away from the boundary $\langle S^z(r)\rangle$ becomes a constant, the
bulk magnetization
\begin{eqnarray}
\label{eq_FT6}
m &=& \lim_{r\rightarrow\infty}\lim_{N\rightarrow\infty}\langle S^z(r)\rangle \nonumber\\
&=& \lim_{r\rightarrow\infty}\frac{\sum_n\langle\Psi_L^0|T(S^z)T^{r-1}|\Psi_R^n\rangle\langle\Psi_L^n|\wt{T}|\Psi_R^0\rangle}
{\Lambda_0^r \langle\Psi_L^0|\wt{T}|\Psi_R^0\rangle} \\
&=& \frac{\langle\Psi_L^0|T(S^z)|\Psi_R^0\rangle}{\Lambda_0} \nonumber  \; .
\end{eqnarray}
To obtain numerically the susceptibility profile $\chi_B(r)$ we calculate
$C(r)$ for small fields $h\sim 10^{-4}$, $10^{-5}$ by using
Eqs.~(\ref{eq_FT5},\ref{eq_FT6}) and then taking the numerical derivative.

\begin{figure} 
\begin{center}
\includegraphics*[width=0.99\columnwidth]{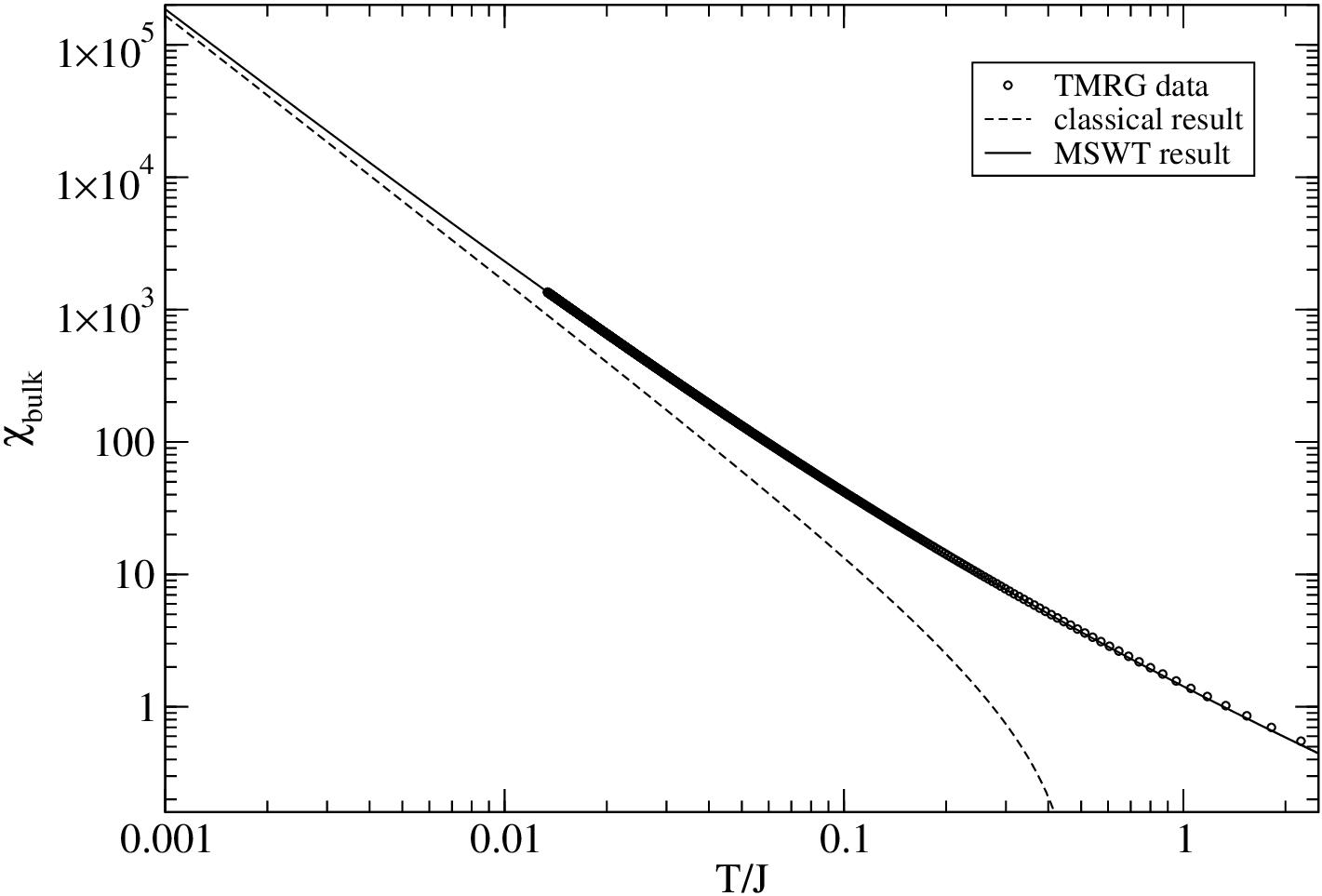} 
\end{center}
\caption{$\chi_{\text{bulk}}$ as a function of
  temperature. The circles denote the numerical data obtained by TMRG, the
  dashed line is the classical result from Eq.~(\ref{eq_i3}) and the solid
  line Takahashi's result (\ref{eq_MSWT11}) obtained by MSWT.}
\label{fig0} 
\end{figure}

Here we want to study the quantum model (\ref{eq_i1}) with $S=1/2$, $J=1$ and
$g=2$. First, we want to test our numerical results by calculating the bulk
susceptibility and comparing with Eq.~(\ref{eq_MSWT11}) which agrees with the
TBA. \cite{TakahashiMSWT} The result is shown in Fig.~\ref{fig0} and the
agreement at low temperatures is excellent.
Note also, that although the leading terms in the low-temperature
expansion for the classical and the quantum model are identical,
extremely low temperatures are necessary to see the classical scaling
for the $S=1/2$ quantum model.

\begin{figure} 
\begin{center}
\includegraphics*[width=0.99\columnwidth]{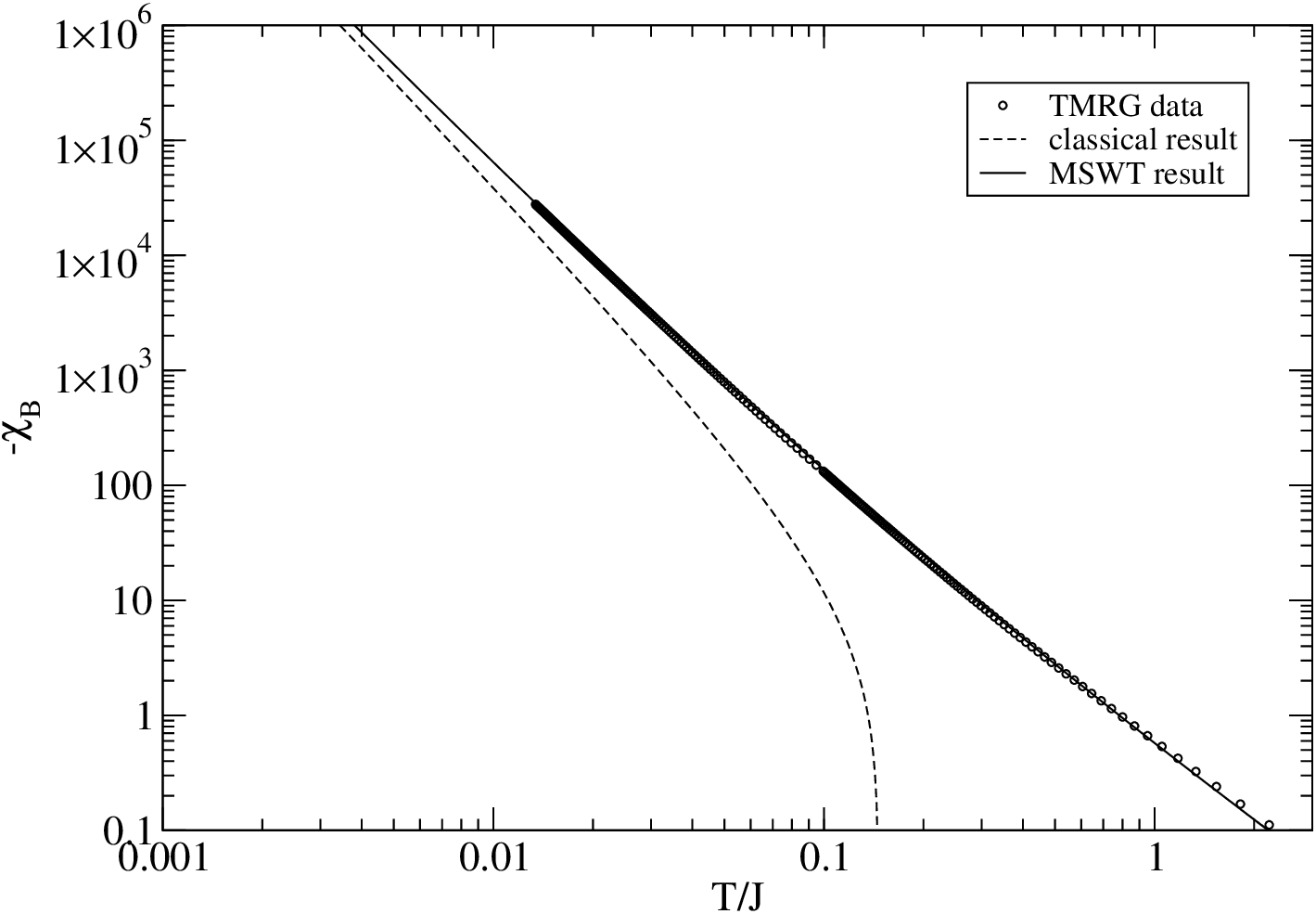} 
\end{center}
\caption{Boundary susceptibility $\chi_{B}$ as a function of
temperature. The circles denote the numerical data obtained by TMRG,
the dashed line is the classical result from Eq.~(\ref{eq_i4}) and the
solid line our result (\ref{eq_MSWT13}) from MSWT.}  
\label{fig1} 
\end{figure}

The boundary susceptibility is shown in Fig.~\ref{fig1} in comparison
to the classical result as well as to formula (\ref{eq_MSWT13})
conjectured for the quantum case. 
The excellent agreement confirms our conjecture for the $S=1/2$
case. As (\ref{eq_MSWT13}) also agrees with the classical result in
the limit $S\rightarrow\infty$ we expect that our result is valid for
all $S$.

\begin{figure} 
\begin{center}
\includegraphics*[width=0.99\columnwidth]{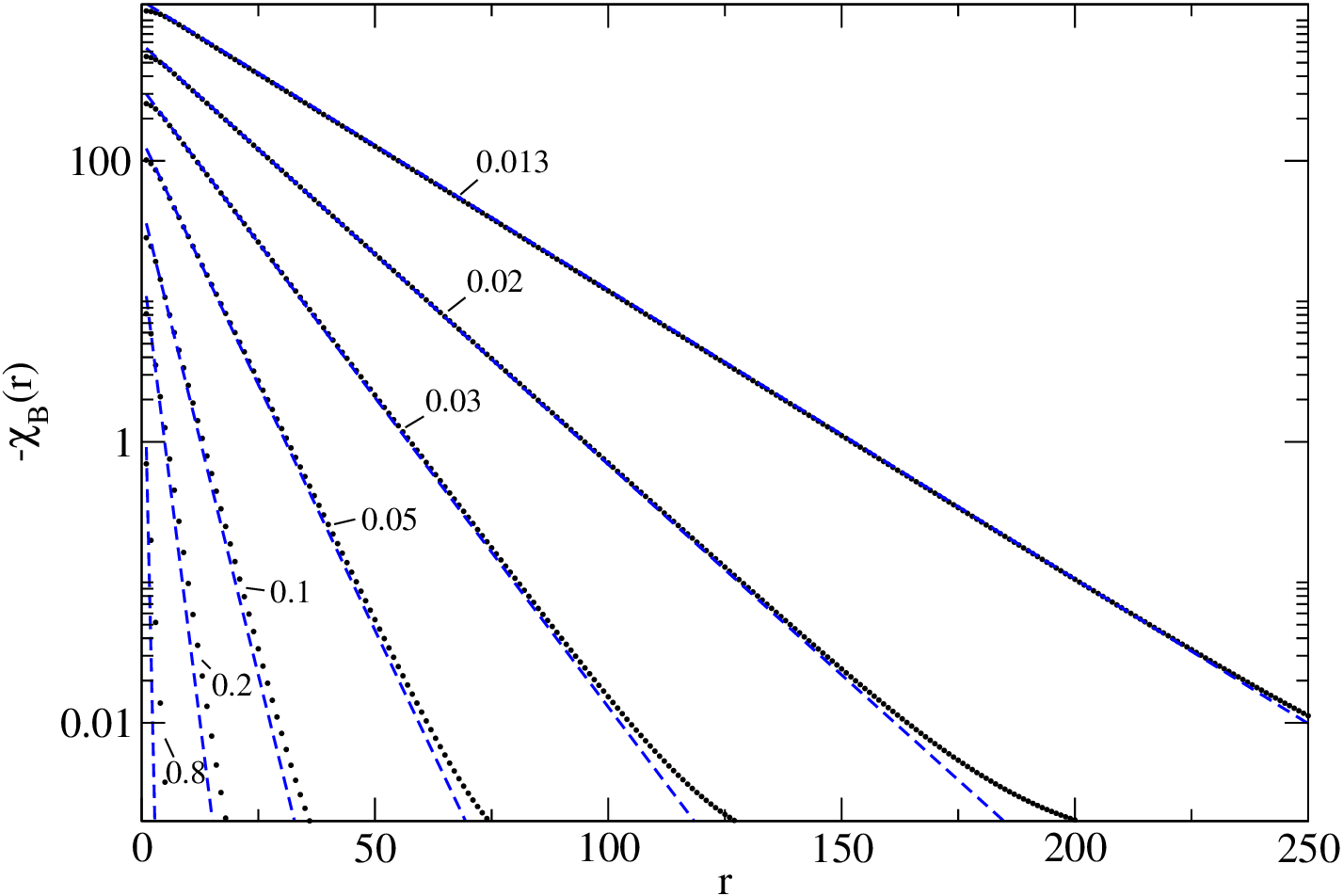}
\end{center} 
\caption{Susceptibility profile $\chi_B(r)$ at a distance $r$ from the boundary 
  for different temperatures $T=0.013,\cdots,0.8$. The dots represent the
  numerical, the dashed lines the theoretical result according to
  Eq.~(\ref{eq_FT7}).}
\label{fig2} 
\end{figure}

Finally, we show in Fig.~\ref{fig2} susceptibility profiles
$\chi_B(r)$ for different temperatures.
As the total boundary susceptibility is given by Eq.~(\ref{eq_FT3})
which, on the other hand, should be equal to (\ref{eq_MSWT13}) we can
even determine an analytic formula for $\chi_B(r)$ and find
\begin{equation}
\label{eq_FT7}
\chi_B(r)=-g^2\frac{\e^{2\sqrt{tv}}-1}{24JS}t^{-1}v^{-2}\e^{-2r\sqrt{tv}} \; .
\end{equation}
This formula is in excellent agreement with our numerical data (see dashed
lines in Fig.~\ref{fig2}). The deviations at large distances $r$ where
$\chi_B(r)$ is small are due to numerical errors. For the fields $h\sim
10^{-5}$ used here, $\chi_B(r)\sim 10^{-2}$ corresponds to a local
magnetization $C(r)\sim 10^{-7}$ which becomes comparable with the accuracy of
the calculation. Note also, that according to Eq.~(\ref{eq_FT7}) the one-point
correlation function $\langle S^z(r)\rangle$ will decay for small magnetic
fields with exactly the same correlation length as the bulk two-point
correlation function $\langle S^z(r)S^z(0)\rangle$. This connection between
one and two-point correlation function has also been found for the
antiferromagnetic $XXZ$ chain. \cite{BortzSirker}
\section{Conclusions}
\label{Conclusion} 
We want to emphasize that the boundary susceptibility is not a finite size
quantity. It is defined as the difference in susceptibilities between a
periodic chain and a chain with OBC in the {\it thermodynamic limit}. In fact,
when we calculated $\chi_{\text{\tiny PBC}}$ and $\chi_{\text{\tiny OBC}}$ in
section \ref{MSWT} we have ignored terms $\sim \exp(-2N\sqrt{tv})$. For a
finite chain with OBC this is a valid approximation if $T/J>1/4N$ and our
results can be directly applied if this condition is fulfilled. At
temperatures $T/J\sim 1/N$, where finite size corrections are sufficiently
small to be ignored, we find a $\sim 25\%$ reduction of the total
susceptibility in the open compared to the periodic system. This effect should
therefore be relevant in susceptibility measurements on systems with
non-magnetic impurities when the temperature $T/J$ becomes comparable to the
concentration of impurities (inverse average chain length).

In this context we want to mention that the low-$T$ behavior of
$\chi_{\text{\tiny PBC}}$ following from Eqs.~(\ref{eq_MSWT9},\ref{eq_MSWT11})
has been observed experimentally. \cite{Takahashi_exp,Takeda_exp} Furthermore,
controlled doping of quasi-one dimensional ferromagnets with both magnetic
\cite{Mukai_exp} and non-magnetic \cite{Narayan_exp} defects is possible. Most
interestingly, susceptibility measurements of diluted two-dimensional
ferromagnets have revealed a one-dimensional behavior at the percolation
threshold, \cite{Okuda_exp} and an unexplained lowering of the susceptibility
under the percolation threshold at low temperatures. It would be certainly
interesting to try to understand these experiments in more detail in the light
of the results presented here.

Finally, we want to address the question at which temperature scale the
crossover from quantum to classical behavior occurs. Clearly, the system
behaves classically at length scales much smaller than the correlation length
$\xi=1/2\sqrt{tv}\approx J_c/T$ where all spins are practically aligned. The
length scale for fluctuations is set by the spin-wave wavelength
$\lambda\sim\sqrt{J_c/TS}$. So we expect classical behavior when
$\lambda\ll\xi$ which is true for all $S$ at sufficiently low temperatures. As
expected, $\lambda$ becomes smaller with increasing $S$ whereas the
correlation length $\xi$ does not change. Therefore the crossover temperature
will increase with the spin quantum number $S$.

In summary, we have used a modified spin-wave theory - where a chemical
potential guarantees zero magnetization at zero magnetic field for any finite
temperature - to calculate the boundary susceptibility $\chi_B$ for the open
spin-$S$ quantum ferromagnetic chain. We found that $\chi_B$ can be expanded
in powers of $\sqrt{T}$ and that the leading term is given by $\chi_B\sim
-1/T^3$ in agreement with the classical result. The quantum corrections to
this classical result are, however, important to obtain a good description
over a wide temperature range. We have verified our formula for the $S=1/2$
case by comparing with numerical data obtained by the density-matrix
renormalization group applied to transfer matrices and have found excellent
agreement. We have been even able to derive an analytic formula for the local
boundary susceptibility $\chi_B(r)$ which we also checked numerically. Most
important, we have shown that $\chi_B$ at low temperatures is ``universal'',
in the sense that it is completely determined by the long-distance asymptotics
of the two-point correlation function $\langle \vec{S}_n\vec{S}_m\rangle$.
\begin{acknowledgments} 
  The authors acknowledge support by the German Research Council ({\it
    Deutsche Forschungsgemeinschaft}) and thank the Westgrid Facility (Canada)
    where the numerical calculations have been performed.
\end{acknowledgments}

\end{document}